\documentclass[preprint,12pt]{elsarticle}

 \usepackage{graphics}

\usepackage{amssymb}

\biboptions{sort&compress}

\journal{Physica C}

\begin{document}

\begin{frontmatter}



\title{Universal intrinsic doping behavior of\\ in-plane dc conductivity for\\ hole-doped high-temperature cuprate superconductors}


\author{Tatsuya Honma}

\address{Department of Physics, Asahikawa Medical University, Asahikawa, Hokkaido 078-8510, Japan}
\ead{honma@asahikawa-med.ac.jp}

\author{Pei Herng Hor}

\address{Department of Physics and Texas Center for Superconductivity, University of Houston,\\ Houston, Texas 77204-5005, U.S.A.}
\ead{phor@uh.edu}

\begin{abstract}
Understanding the normal state transport properties in hole-doped high-temperature cuprate superconductors (HTCSs) is a challenging task which has been widely believed to be one of the key steps toward revealing the pairing mechanism of high-temperature superconductivity. Here, we present a true intrinsic and universal doping dependence of $in$-$plane$ dc conductivity for all underdoped HTCSs. The doping dependence of $in$-$plane$ dc conductivity normalized to that at optimal doping can be represented by a simple exponential formula. The doping behavior of the square of the nodal Fermi velocity derived by the high-resolution laser-based angle-resolved photoemission spectroscopy in the superconducting state  follows reasonably well the universal intrinsic doping behavior. Our findings suggest a commonality of the low-energy quasiparticles both in the normal and superconducting states that place a true universal and stringent constraint on the mechanism of high-temperature superconductivity for HTCSs.
\end{abstract}


\end{frontmatter}


\section{Introduction}
\label{}

Transport measurement in solids, such as dc conductivity and thermoelectric power, probes the low-energy quasiparticles with an unprecedented resolution. In hole-doped high-temperature cuprate superconductors (HTCSs), the normal state transport properties exhibit unusually strong temperature ($T$) dependence which causes substantial difficulties in interpreting the transport data. In contrast, the doping-dependent studies have the distinct advantages of directly studying the evolution of the electronic states upon hole-doping without the complications due to the variation of $T$. If a dopant-induced property depends on hole-doping alone, regardless of dopant and material, then its doping dependence will be the true intrinsic property and warrant serious consideration. The key to reveal a genuine doping dependence is the establishment of a universal hole-scale for measuring the doped-hole contents per CuO$_2$ plane ($P_{pl}$) in HTCS, irrespective of dopant and material. We proposed a hole-scale to serve this purpose \cite{hon04}. In this paper, we re-examined accumulated data of $in$-$plane$ dc resistivity based on the hole-scale and revealed, for the first time, the genuine doping dependence of $in$-$plane$ dc conductivity for HTCSs. 

We constructed the hole-scale based on thermoelectric power at 290 K ($S^{290}$) \cite{hon04}. $P_{pl}$ of La$_{2-x}$Sr$_x$CuO$_4$ (SrD-La214) and Y$_{1-x}$Ca$_x$Ba$_2$Cu$_3$O$_6$ (CaD-Y1236) with no excess oxygen can be uniquely determined from the doped-cation contents alone \cite{hon04}. We found that the $S^{290}$ versus $x/2$ (= $P_{pl}$) curve in CaD-Y1236 is consistent with the $S^{290}$ versus $x$ (= $P_{pl}$) curve in SrD-La214, and then we defined a universal relation of $P_{pl} = -ln(S^{290}/392)/19.7$ \ ($0.01 \le P_{pl} < 0.21$) and $P_{pl} = (40.47-S^{290})/163.4$ ($0.21 \le P_{pl} < 0.34$), where the unit of $S^{290}$ is the microvolt per Kelvin ($\mu$V/K) \cite{hon04,hon08}. We call this relation the universal intrinsic hole-scale for HTCSs or \textquotedblleft $P_{pl}$-scale\textquotedblright. We showed that the doped-hole concentration determined by $P_{pl}$-scale is consistent with that by the other techniques \cite{hon08}. The proposed $P_{pl}$-scale has three distinct merits: it can be used for various HTCSs with equivalent CuO$_2$ planes, such as pure cation-, pure anion-, and cation/anion co-doped HTCSs; it can be easily measured at room temperature and it can be used for different sample forms whether if it is single crystal, thin film or polycrystal \cite{hon04,hon08}.

\begin{figure*}
\includegraphics[scale=0.6]{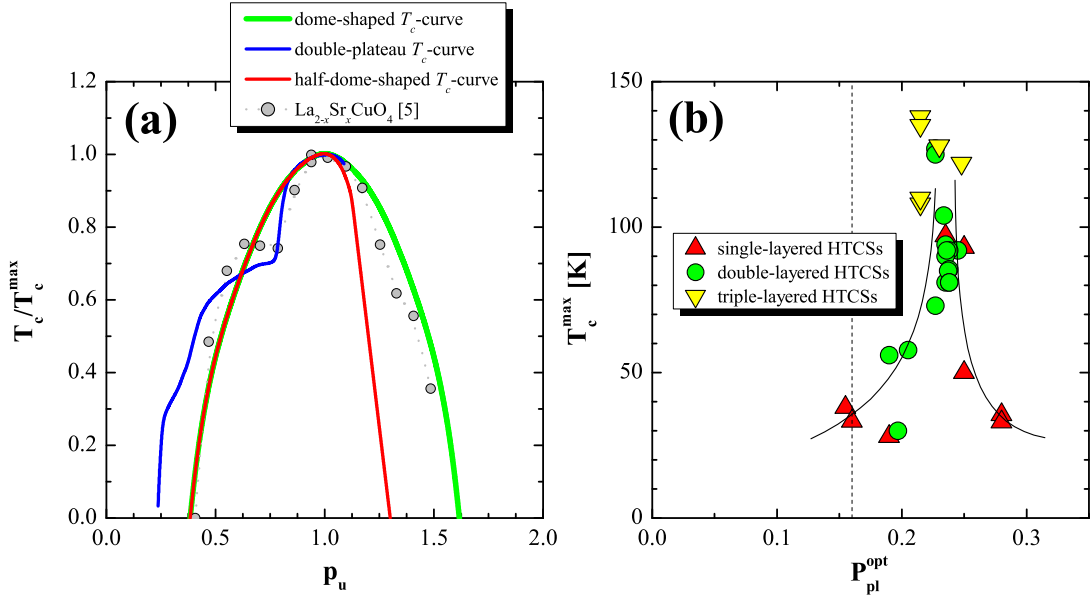}
\caption{\label{fig1}(a) $T_c/T_c^{max}$ of almost all HTCSs as a function of $p_u$ = $P_{pl}/P_{pl}^{opt}$. The green, blue and red curves show a well known dome-shaped $T_c$-curve \cite{pre91}, a double plateau $T_c$-curve \cite{hon07}, and an asymmetric half-dome-shaped $T_c$-curve \cite{hon08}, respectively. For comparison, the observed values of $T_c$ in La$_{2-x}$Sr$_x$CuO$_4$ are plotted \cite{rad94}. (b) $T_c^{max}$ as a function of $P_{pl}^{opt}$. The plotted data come from Table VII in Ref.\ \cite{hon08}. $P_{pl}^{opt}$ (= 0.16) in La$_{2-x}$Sr$_x$CuO$_4$ is close to a lower limit of $P_{pl}^{opt}$ for HTCSs. The majority of HTCSs has $P_{pl}^{opt}$ located between 0.22 and 0.25. The dashed line shows the relation based on the popular scale of Ref.\ \cite{pre91}. On the scale, all the values of $T_c^{max}$ must lie on the dashed line. Two solid lines are guide to the eyes.}
\end{figure*}

Until now, the doping dependence of $T_c$ for HTCSs has been commonly believed to follow a symmetrical dome-shaped $T_c$-curve \cite{pre91}. This was established on the basis of the $T_c$ versus doped-cation contents in SrD-La214 \cite{tak89} and the doping dependence of $T_c$ for other HTCSs was just assumed to follow the same doping dependence as that of SrD-La214. Strictly speaking even SrD-La214 does not follow the dome-shaped $T_c$-curve \cite{rad94}. We have examined the doping dependence of $T_c$ for other HTCSs by $P_{pl}$-scale, and found that they follow an asymmetric half-dome-shaped $T_c$-curve \cite{hon08}. Further, although the value of optimal doped-hole concentration ($P_{pl}^{opt}$) for SrD-La214 is 0.16, the majority of HTCSs has $P_{pl}^{opt}$ located between 0.22 and 0.25 as shown in Fig.\ \ref{fig1}(b) \cite{hon08}. Only YBa$_2$Cu$_3$O$_{6+\delta}$ (OD-Y123) follows a double-plateau $T_c$-curve with $P_{pl}^{opt}$ = 0.25 \cite{hon07}. Three types of the doping dependence of $T_c$ are summarized in Fig.\ \ref{fig1}(a). For comparison, we plot $T_c/T_c^{max}$ as a function of $P_{pl}/P_{pl}^{opt}$. $T_c^{max}$ and $P_{pl}^{opt}$ are the maximum $T_c$ and the corresponding $P_{pl}$ on the $T_c$ versus $P_{pl}$ curve for a HTCS material, respectively. We called the value of $P_{pl}/P_{pl}^{opt}$ a unified doped-hole concentration ($p_u$) \cite{hon08}. Even if we have no $S^{290}$ value of a sample, we can determine the value of $p_u$ from its $T_c$, according to the corresponding $T_c/T_c^{max}$-curve in Fig.\ \ref{fig1}(a). This secondary procedure to determine $p_u$ is very useful, because the research reports on HTCSs had almost always included the value of $T_c$. We call this second scale the \textquotedblleft $p_u$-scale\textquotedblright. 

While there are many $T$-dependent studies of $in$-$plane$ dc conductivity \cite{kom05,kom03,xu00,nak93,abe99,hay91,nak92,wak04,koh03,dai01,nag08,yak95,noj03,seg03,tak94,lee05,coo00,bab99,sem01,zve03,wan01,and00,men09,ama04,kon06,fuj02,wat97,fuj95,yam07,fuj02b}, there are relatively few studies that qualitatively compare the $T$ dependence on the different HTCS materials near the optimal doping level or $P_{pl}^{opt}$ \cite{iye92}. There is no doping-dependent study that quantitatively compares $in$-$plane$ dc conductivity of many different HTCSs over a wide doping range. By converting the doped-hole concentration in the literature to that based on $P_{pl}$-scale, we revealed an intrinsic doping dependence of various physical properties in HTCSs \cite{hon04,hon08,hon07,hon06}. In particular, we uncovered a universal electronic phase diagram (UEPD) \cite{hon08}, which is constructed on $p_u$. In this paper, using $p_u$, we have analyzed the doping dependence of $in$-$plane$ dc conductivity, including dc conductivity along the $a$-axis ($\sigma_a$), dc conductivity along the $b$-axis ($\sigma_b$), and $in$-$plane$ dc conductivity of the twined crystal ($\sigma_{ab}$), of 13 HTCS materials, reported in 30 published papers that covers almost all major HTCS materials reported up to date \cite{kom05,kom03,xu00,nak93,abe99,hay91,nak92,wak04,koh03,dai01,nag08,yak95,noj03,seg03,tak94,lee05,coo00,bab99,sem01,zve03,wan01,and00,men09,ama04,kon06,fuj02,wat97,fuj95,yam07,fuj02b}. We unravel a universal intrinsic doping dependence of $in$-$plane$ dc conductivity that is valid not only, surprisingly, over a wide temperature range that covers physically important pseudogap regime, but also valid for the various structures ranging from structure that based on the simple single CuO$_2$ layer to that composing of complex multiple CuO$_2$ layers.

\begin{table}[b]
\caption{\label{tab:table1} The information for the data plotted in Figs.\ \ref{fig2} and \ \ref{fig3}. The value of $P_{pl}^{opt}$ is coming from Refs.\ \cite{hon04,hon08,hon07,hon06}.}
\begin{tabular}{clccl} 

\\
\hline
Fig. & Materials                                            & scale & ${P_{pl}^{opt}}$ & Ref(s). \\ 

\hline
2    & La$_{2-x}$Sr$_x$CuO$_4$                              & $P_{pl}$   & 0.16  & \cite{kom05,kom03,xu00,nak93} \\
     & La$_{2-x}$Ba$_x$CuO$_4$                              & $P_{pl}$   & 0.16  & \cite{abe99,hay91} \\
     & La$_{1.6-x}$Nd$_{0.4}$Sr$_x$CuO$_4$                  & $P_{pl}$   & 0.16  & \cite{nak92} \\
     & Ca$_{2-x}$Na$_x$CuO$_2$Cl$_2$                        & $P_{pl}$   & 0.19  & \cite{wak04,koh03} \\
     & (Hg$_{0.8}$Cu$_{0.2}$)Ba$_2$CuO$_{4+\delta}$         & $P_{pl}$   & 0.235 & \cite{dai01} \\
     & Y$_{1-x}$Ca$_{x}$Ba$_2$Cu$_3$O$_{6+\delta}$          & $p_u$      &   NA   & \cite{nag08} \\
     & Y$_{1-x}$Ca$_{x}$Ba$_2$Cu$_3$O$_{6+\delta}$          & $P_{pl}$   & 0.235 & \cite{yak95} \\
     & Ca$_{0.5}$La$_{1.25}$Ba$_{1.25}$Cu$_3$O$_y$          & $p_u$      &  NA   & \cite{noj03} \\
\hline
3    & YBa$_2$Cu$_3$O$_{6+\delta}$                          & $P_{pl}$   & 0.25  & \cite{seg03,wan01} \\
     & YBa$_2$Cu$_3$O$_{6+\delta}$                          & $p_u$      &  NA   & \cite{nag08,tak94,lee05,coo00,bab99,sem01,zve03} \\
\hline
\end{tabular} 
\end{table}

\section{Analysis}

In analyzing the $in$-$plane$ dc conductivity, we selected the single crystal data with either $S^{290}$ or $T_c$ among the published data. We use two previously mentioned methods to extract $p_u$: as the first method, $P_{pl}$ is determined from the value of $S^{290}$ by using $P_{pl}$-scale and the value of $p_u$ is calculated by dividing the $P_{pl}$ value by the corresponding $P_{pl}^{opt}$ value. This is more reliable \cite{hon04,hon08}. As the second method, $p_u$ is determined from the value of $T_c/T_c^{max}$ by comparing it with the corresponding $T_c/T_c^{max}$-curve as shown in Fig.\ \ref{fig1}(a). For the accurate analysis, we always selected the paper that also reports the value of $S^{290}$ first and used the data with the value of $T_c$ second. However, this method restricts the available data. In the so-called \textquotedblleft 214 system\textquotedblright, such as SrD-La214, La$_{2-x}$Ba$_x$CuO$_4$ and La$_{1.6-x}$Nd$_{0.4}$Sr$_x$CuO$_4$, the doped-cation contents directly give us $P_{pl}$. Since the value of $P_{pl}^{opt}$ is 0.16, the value of $p_u$ is determined by dividing each doped-cation contents by 0.16. For OD-Y123, the value of $p_u$ was estimated from the double plateau $T_c$/$T_c^{max}$-curve in Fig.\ \ref{fig1}(a). For other HTCSs, the value of $p_u$ was determined by comparing the $T_c$/$T_c^{max}$ value with the asymmetrical half-dome-shaped $T_c$-curve as shown in Fig.\ \ref{fig1}(a). Off course, whenever the materials have the $S^{290}$ value, the value of $P_{pl}$ was always directly determined by using $P_{pl}$-scale, and then the value of $p_u$ is calculated by dividing the $P_{pl}$ value by the corresponding $P_{pl}^{opt}$ value in Table VII of Ref.\ \cite{hon08}.

\section{Results and Discussion}

\begin{figure}
\includegraphics[scale=0.35]{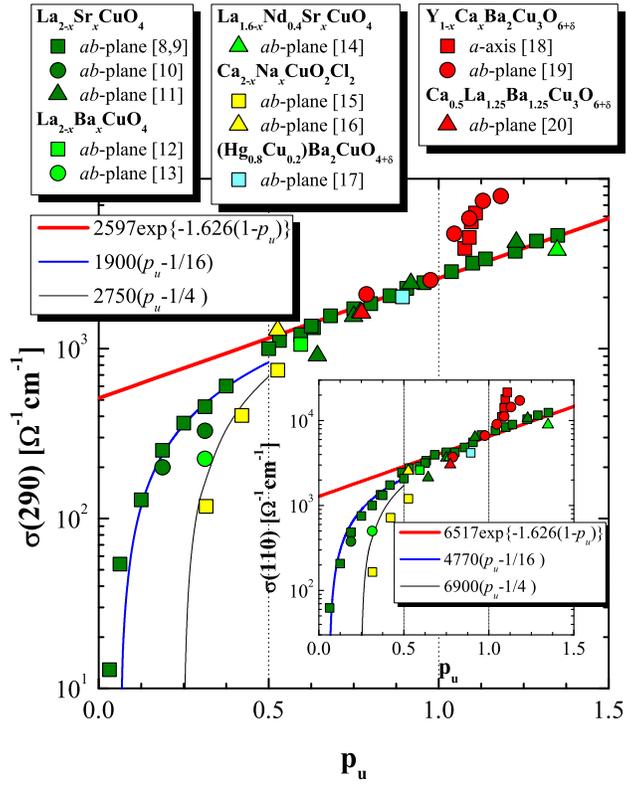}
\caption{\label{fig2}$\sigma(290)$ versus $p_u$ for the major single-, and double-layered HTCSs with cation and/or anion co-doped. Inset: $\sigma(110)$ versus $p_u$. The plotted data are summarized in Table~\ref{tab:table1}.}
\end{figure}

\begin{figure}
\includegraphics[scale=0.35]{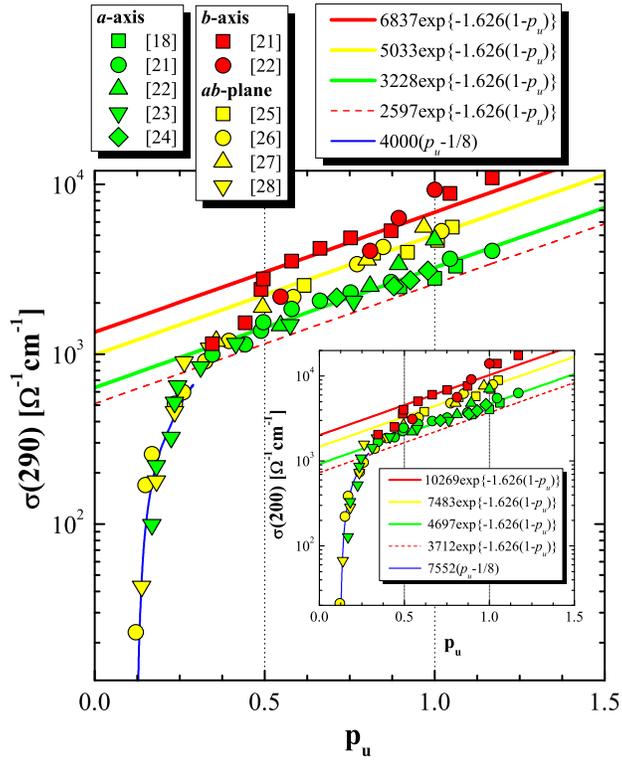}
\caption{\label{fig3}$\sigma(290)$ versus $p_u$ for YBa$_2$Cu$_3$O$_{6+\delta}$. The inset shows $\sigma(200)$ versus $p_u$. The plotted data are summarized in Table~\ref{tab:table1}. }
\end{figure}

\begin{table*}
\caption{\label{tab:table2} The information for the data plotted in Fig.\ \ref{fig4}. The value of $P_{pl}^{opt}$ is coming from Refs.\ \cite{hon04,hon08,hon07,hon06}. It does not need for $p_u$-scale. The values of $\sigma^{opt}(290)$ and $\sigma^{opt}(130)$ are scaling factor.  }
\begin{tabular}{lcccccc}
\\
\hline
 Materials & scale     & ${P_{pl}^{opt}}$ & $\sigma^{opt}$(290)    & $\sigma^{opt}$(130)    & Ref.   \\
           &                &                  & [$\Omega^{-1}cm^{-1}$] & [$\Omega^{-1}cm^{-1}$] &           \cr
\hline           
Bi$_2$Sr$_{2-x}$La$_x$CuO$_{6+\delta}$               & $P_{pl}$   & 0.28  & 2.03$\times10^3$ & 4.30$\times10^3$ & \cite{and00} \\
Bi$_2$Sr$_{2-x}$La$_x$CuO$_{6+\delta}$               & $p_u$      &  NA   & 2.09$\times10^3$ & 2.94$\times10^3$ & \cite{men09} \\
Bi$_{1.7}$Pb$_{0.3}$Sr$_{2-x}$La$_x$CuO$_{6+\delta}$ & $p_u$      &  NA   & 2.05$\times10^3$ & 4.30$\times10^3$ & \cite{ama04} \\
(Bi,Pb)$_2$(Sr,La)$_2$CuO$_{6+\delta}$               & $p_u$      &  NA   & 1.78$\times10^3$ & 2.94$\times10^3$ & \cite{kon06} \\
Bi$_2$Sr$_2$CaCu$_2$O$_{8+\delta}$ ($a$-axis)        & $P_{pl}$   & 0.238 & 1.54$\times10^3$ & 3.24$\times10^3$ & \cite{fuj02} \\
Bi$_2$Sr$_2$CaCu$_2$O$_{8+\delta}$ ($b$-axis)        & $P_{pl}$   & 0.238 & 1.42$\times10^3$ & 2.78$\times10^3$ & \cite{fuj02} \\
Bi$_2$Sr$_2$CaCu$_2$O$_{8+\delta}$                   & $p_u$      &  NA   & 2.03$\times10^3$ & 4.00$\times10^3$ & \cite{wat97} \\
Bi$_2$Sr$_2$CaCu$_2$O$_{8+\delta}$                   & $p_u$      &  NA   & 1.65$\times10^3$ & 2.48$\times10^3$ & \cite{fuj95} \\
Bi$_2$Sr$_2$CaCu$_2$O$_{8+\delta}$                   & $p_u$      &  NA   & 2.92$\times10^3$ & 5.96$\times10^3$ & \cite{yam07} \\
Bi$_2$Sr$_2$Ca$_2$Cu$_3$O$_{10+\delta}$              & $P_{pl}$   & 0.215 & 2.09$\times10^3$ & 6.33$\times10^3$ & \cite{fuj02b} \\
Bi$_2$Sr$_2$Ca$_2$Cu$_3$O$_{10+\delta}$              & $p_u$      &  NA   & 2.09$\times10^3$ & 6.33$\times10^3$ & \cite{fuj02b} \\   
\hline 
\end{tabular}
\end{table*}

First of all, for HTCSs without the doping-induced extend structural features like CuO chain of OD-Y123, we demonstrate that $in$-$plane$ dc conductivity at a fixed temperature depends on $p_u$ alone. In Fig.\ \ref{fig2}, we plot $in$-$plane$ dc conductivity at 290 K ($\sigma(290)$) for major single-, and double-layered HTCSs with cation and/or anion co-doping as a function of $p_u$ on the semi-logarithmic plot. Since $\log \sigma(290)$ linearly increases with doping in the underdoped range of $0.5 \le p_u < 1$, $\sigma(290)$ seems to depend on $p_u$ exponentially. For $p_u < 0.5$, $\sigma(290)$ rapidly deviates downward from the linear dependence of $\log \sigma(290)$ with undoping. The downward deviation tends to follow a linear $p_u$ dependence originating at either $p_u$ = 1/16 or 1/4. In the overdoped range of $p_u > 1$, the doping dependence of $\sigma(290)$ can be roughly separated into two types of behaviors. One keeps on following the same behavior in the underdoped range and the other deviates upward from it at $p_u$ = 1. The doping dependence at 110 K is shown in the inset of Fig.\ \ref{fig2}. We observe the identical behavior as that at 290 K. Same behavior was observed in the temperature range from 110 K to 290 K. Below 110 K, it becomes increasingly more difficult to reliably pin down the systematic behavior of $in$-$plane$ dc conductivity, since many plotted HTCS materials are superconducting below 110 K. Finally, from Fig.\ \ref{fig2}, when we assume a simple exponential function as the doping dependence at $0.5 \le p_u < 1$, we can express the $in$-$plane$ dc conductivity at $T$ and $p_u$ as 
\begin{eqnarray}
\sigma(p_u,T)/\sigma^{opt}(T) =
 \left.
 \begin{array}{ll}
 \exp\{-1.626(1-p_u)\}.  & \label{form1}
\end{array}
\right.
 \\ (0.5 \le p_u < 1). \nonumber 
\end{eqnarray}
Here, $\sigma^{opt}$($T$) is the $T$ dependence of conductivity at $p_u$ = 1 or optimal conductivity. The values of $\sigma^{opt}$(290) and $\sigma^{opt}$(110) are 2597 [$\Omega^{-1}cm^{-1}$] and 6517 [$\Omega^{-1}cm^{-1}$], respectively. However, since the change in $\sigma$(290) is less than one order of magnitude in the small doping range of $0.5 \le p_u < 1$, the polynominal fit may work also. In fact, the polynominal fit up to fourth order worked equally well as the formula (\ref{form1}) at $0.5 \le p_u < 1$. Since the fitting result has no essential difference, hereafter, we will use the formula (\ref{form1}) as a matter of convenience.

Next, we demonstrate that the doping dependence of in-plane conductivity for $0.5 \le p_u < 1$ is not modified by the doping-induced extended structural features. One representative doping-induced structural features of HTCSs is the CuO chain structure in OD-Y123. Figure\ \ref{fig3} shows $\sigma$(290) of OD-Y123 as a function of $p_u$ on a semi-logarithmic plot. We see that, for $p_u \ge 0.5$, although $\sigma_a$ and $\sigma_b$ are higher than that of co-doped HTCSs in Fig.\ \ref{fig2}, both $\sigma_a$ and $\sigma_b$ still follow formula (\ref{form1}). The $\sigma^{opt}$(290) value is 3228 [$\Omega^{-1}cm^{-1}$] for $\sigma_a$ and 6837 [$\Omega^{-1}cm^{-1}$] for $\sigma_b$. The yellow solid line is the curve fitted to formula (\ref{form1}) with $\sigma^{opt}$(290) = 5033 [$\Omega^{-1}cm^{-1}$] which is just the average of $\sigma^{opt}$(290) for $\sigma_a$ and that for $\sigma_b$. Almost all $\sigma_{ab}$ tend to lie on this averaged curve. At $p_u$ $\sim$ 0.5, $\sigma_b$($p_u$) jumps down with undoping and merges into $\sigma_a$($p_u$). $\sigma_a$ follows formula (\ref{form1}) for $0.3 < p_u < 1.15$. For $p_u$ $<$ 0.25, all conductivities finally follow one linear $p_u$ dependence originated at $p_u$ = 1/8. $\sigma$(200) of OD-Y123 is plotted as a function of $p_u$ in the inset of Fig.\ \ref{fig3}. We observe the identical behavior as that at 290 K. Similar behavior was observed also in the temperature range from 110 K to 290 K.

\begin{figure}
\includegraphics[scale=0.35]{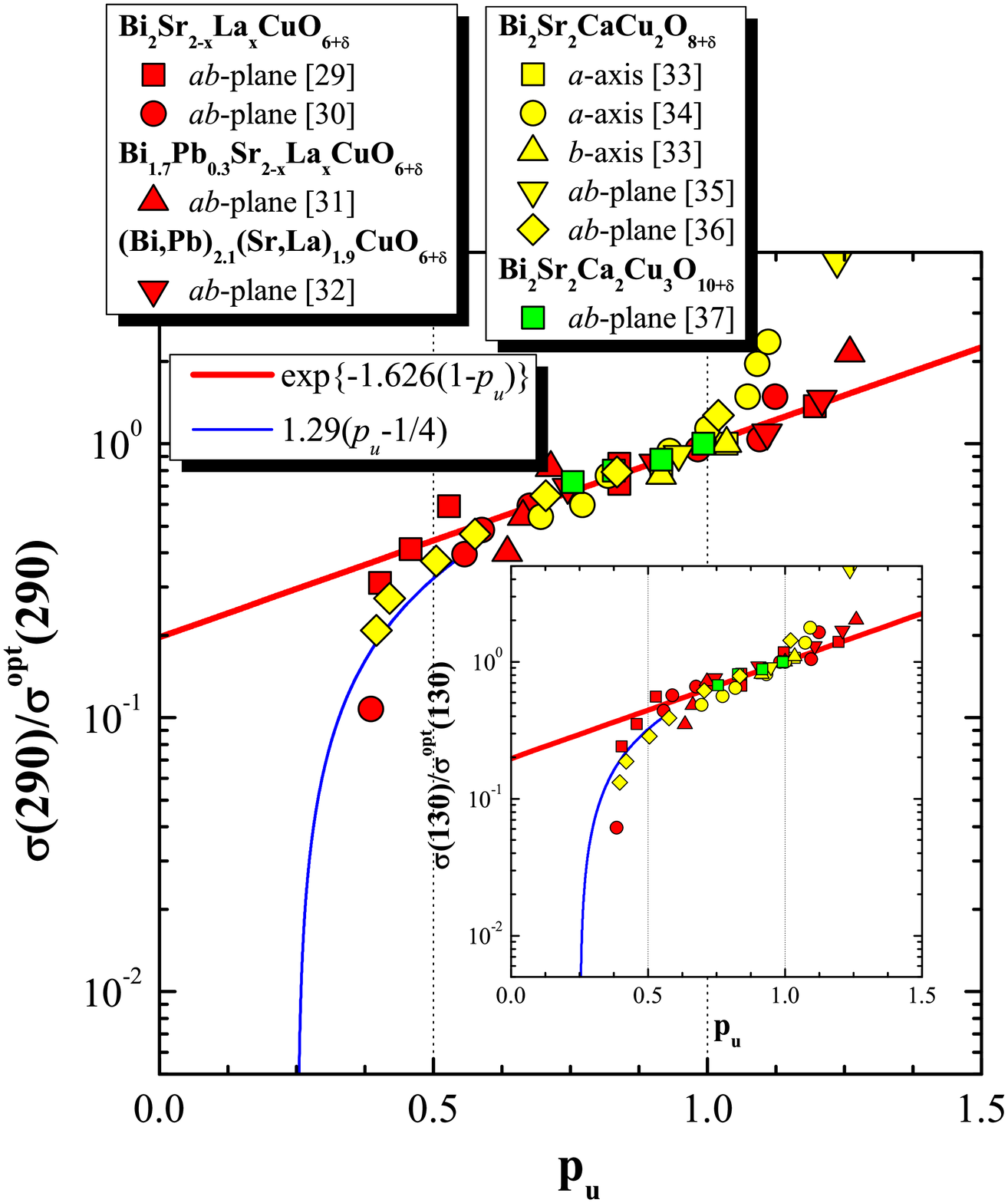}
\caption{\label{fig4}$\sigma(290)/\sigma^{opt}(290)$ versus $p_u$ for the Bi-based family. The inset shows $\sigma(130)/\sigma^{opt}(130)$ versus $p_u$. The plotted data are summarized in Table~\ref{tab:table2}.}
\end{figure}

As another example of the doping-induced extended structural features, in Fig.\ \ref{fig4}, we plot $\sigma(290)/\sigma^{opt}(290)$ for the Bi-based family on a semi-logarithmic plot. For example, Bi$_2$Sr$_2$CaCu$_2$O$_{8+\delta}$ (OD-Bi2212) has a well-known incommensurate structural modulation along the $b$-axis \cite{yam90}. The incommensurate structural modulation reduced $\sigma_b$ \cite{hon91}. Since we want to study the doping dependence alone, we plot the $\sigma(290)$ normalized by the optimal conductivity. Again, we see that $\sigma(290)/\sigma^{opt}(290)$ follows formula (\ref{form1}) for $p_u \ge 0.5$. For $p_u < 0.5$, $\sigma(290)/\sigma^{opt}(290)$ also deviates downward from the exponential dependence with undoping and tends to follow a linear $p_u$ dependence originating at $p_u$ = 1/4. Same result is observed also at 130 K as shown in the inset of Fig.\ \ref{fig4}. Further, identical behavior was observed also in the temperature range from 110 K to 290 K (from 130 K to 290 K for Bi$_2$Sr$_2$Ca$_2$Cu$_3$O$_{8+\delta}$ with $T_c^{max}$ = 108 K). This leads us to conclude that the Bi-based family, irrespective of its structural modulation, follows formula (\ref{form1}). Therefore we conclude that the doping dependence of in-plane conductivity for $0.5 \le p_u < 1$ is not influenced by the doping-induced extended structural features, such as the CuO chain and modulation structure, and they only affect the magnitude of dc conductivity through a doping-independent scaling factor of $\sigma^{opt}$. 

From Figs.\ \ref{fig2} -\ \ref{fig4}, it is indeed very surprising, in light of the many different types of block layers, oxygen coordination, the doping-induced extended structural features, and number of CuO$_2$ layers involved in the crystals, that the doping and temperature dependence of $in$-$plane$ dc conductivity for $0.5 \le p_u < 1$ is universal, and we will call this doping dependence the \textquotedblleft $universal$ $intrinsic$ $doping$ $behavior$\textquotedblright for the in-plane conductivity of HTCSs. Note that the $universal$ $intrinsic$ $doping$ $behavior$ is derived as a function of $p_u$, which is defined as $P_{pl}/P_{pl}^{opt}$. This trend strongly implies, in contrast to the general belief, that $P_{pl}^{opt}$ is not simply an accidental crossover point between pseudogap and superconductivity. The electronic state at $P_{pl}^{opt}$ plays a peculiar and prominent role such that both the normal and superconducting properties are scaled to those at $P_{pl}^{opt}$. The electronic state at $P_{pl}^{opt}$ is the ultimate building block for high-$T_c$ superconductivity. Further, by comparing with the doping dependence of $T_c$ in Fig.\ \ref{fig1}(a), the $p_u$ range for the $universal$ $intrinsic$ $doping$ $behavior$ is corresponding to that for the underdoped superconducting phase. The $universal$ $intrinsic$ $doping$ $behavior$ may be intimately related to the appearance of high-$T_c$ superconductivity.

\begin{figure}
\includegraphics[scale=0.35]{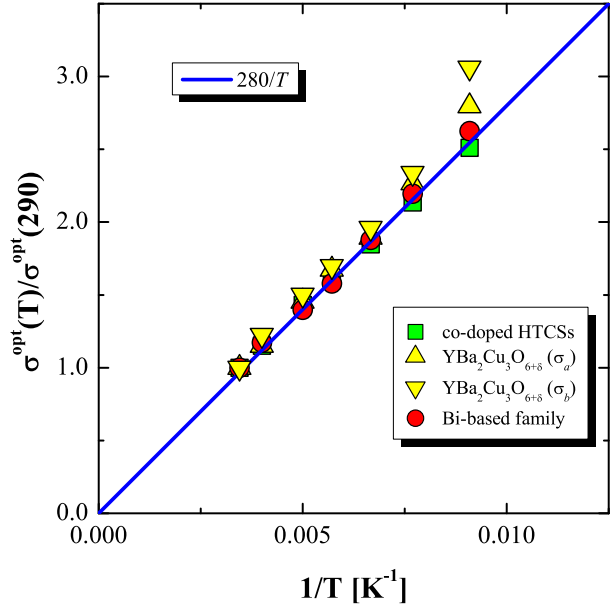}
\caption{\label{fig5}$\sigma^{opt}(T)/\sigma^{opt}(290)$ versus 1/$T$ for cation and anion co-doped HTCSs, YBa$_2$Cu$_3$O$_{6+\delta}$ and the Bi-based family.}
\end{figure}

In Fig.\ \ref{fig5}, we plot the value of the optimal conductivity normalized by the value at 290 K ($\sigma^{opt}(T)/\sigma^{opt}(290)$) as a function of the inverse temperature (1/$T$). There is no difference in cation and anion co-doped HTCSs, OD-Y123 and the Bi-based family. The value of $\sigma^{opt}(T)/\sigma^{opt}(290)$ is clearly proportional to 1/$T$. This is consistent with the linear $T$ dependence of $in$-$plane$ resistivity observed in optimally doped HTCSs. $\sigma^{opt}(T)/\sigma^{opt}(290)$ is represented by following formula:
\begin{eqnarray}
 \sigma^{opt}(T)/\sigma^{opt}(290) =
 \left.
 \begin{array}{ll}
 280/T & \label{form2}
\end{array}
\right.
 \\ (110 \ \mathrm{K} \le T \le 290 \ \mathrm{K}). \nonumber 
\end{eqnarray}

\begin{figure}
\includegraphics[scale=0.35]{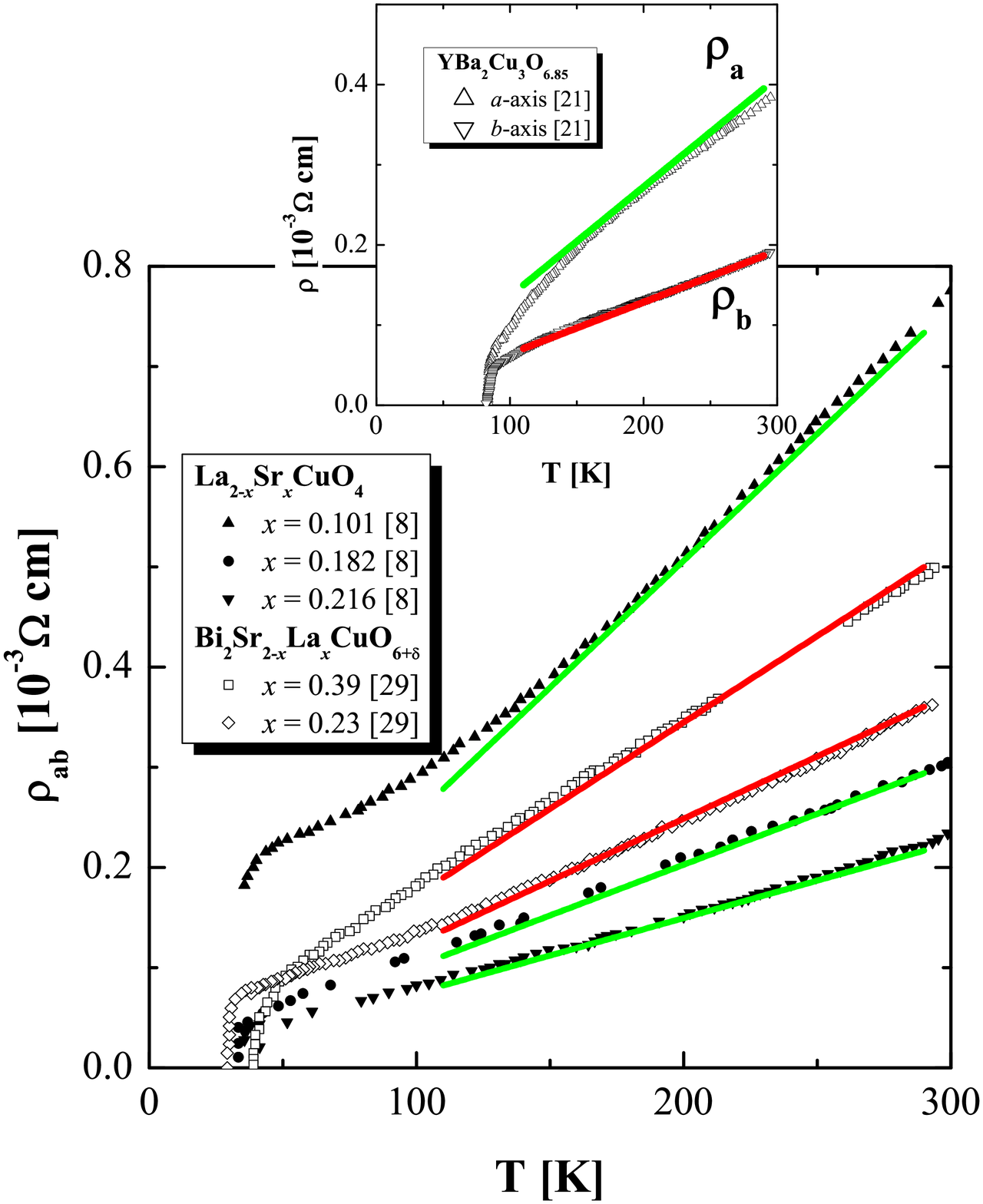}
\caption{\label{fig6}Comparison of simulated $T$ dependence of $in$-$plane$ resistivity using formula (\ref{form3}) to the observed data for typical HTCSs. The values of $\sigma^{opt}(290)$ are 2597 [$\Omega^{-1}cm^{-1}$] for La$_{2-x}$Sr$_x$CuO$_4$, 2028 [$\Omega^{-1}cm^{-1}$] for Bi$_2$Sr$_{2-x}$La$_x$CuO$_{6+\delta}$, 3228 [$\Omega^{-1}cm^{-1}$] for the $a$-axis of YBa$_2$Cu$_3$O$_{6+\delta}$, and 6837 [$\Omega^{-1}cm^{-1}$] for the $b$-axis of YBa$_2$Cu$_3$O$_{6+\delta}$, respectively.}
\end{figure}

Now if we plug formula (\ref{form2}) into (\ref{form1}) we arrive at the following final universal expression of ${\sigma(p_u,T)}$ for HTCSs:
\begin{eqnarray}
 \sigma(p_u,T)/ \sigma^{opt}(290) =
 \left.
 \begin{array}{ll}
 (280/T) \exp \{ -1.626(1-p_u) \} & \label{form3}
 \end{array}
\right.
 \\ (0.5 \le p_u < 1 \ \mathrm{and} \ 110 \ \mathrm{K} \le T \le 290 \ \mathrm{K}). \nonumber 
\end{eqnarray}

In the temperature range of 110 K $\le T \le$ 290 K, irrespective of the dopant type and structural details, $in$-$lane$ dc conductivity of HTCSs consists of a product of the inverse temperature and a temperature-independent exponential term, $\exp\{-1.626(1-p_u)\}$. The value of $\sigma^{opt}(290)$ is the only material-dependent term in formula (\ref{form3}).

\begin{table*}
\caption{\label{tab:table3} The information for the data plotted in Fig.\ \ref{fig7}. The value of $P_{pl}^{opt}$ is coming from Refs.\ \cite{hon04,hon08,hon07,hon06}. The values of $\sigma^{opt}$(290) and $\sigma^{opt}$(150) are scaling factor. Samples coming form Refs.\ \ \cite{kom05} - \cite{nak93} are single crystal. The other samples are polycrystal. }
\begin{tabular}{lcccccl}
\\
\hline
 Materials                            & scale & ${P_{pl}^{opt}}$ & $\sigma^{opt}$(290)    & $\sigma^{opt}(150)$    & $\sigma^{opt}(150)/\sigma^{opt}(290)$ & Ref(s). \\
                                      &            &                  & [$\Omega^{-1}cm^{-1}$] & [$\Omega^{-1}cm^{-1}$] &                                         &         \cr
\hline
La$_{2-x}$Sr$_x$CuO$_4$               & Sr-content & 0.16  & 2.60$\times10^3$ & 4.80$\times10^3$ & 1.85 & \cite{kom05,kom03,xu00,nak93} \\ 
La$_{2-x}$Sr$_x$CuO$_4$               & Sr-content & 0.16  & 990              & 1.98$\times10^3$ & 2.00 & \cite{tak89} \\ 
La$_{2-x}$Sr$_x$CuO$_4$               & Sr-content & 0.16  & 652              & 1.21$\times10^3$ & 1.86 & \cite{nak94} \\ 
YBa$_2$Cu$_{3-x}$Co$_x$O$_{7-\delta}$ & $P_{pl}$   & 0.25  & 951              & 1.63$\times10^3$ & 1.71 & \cite{fis96} \\ 
CaLaBaCu$_3$O$_{6+\delta}$            & $P_{pl}$   & 0.235 & 168              & 236              & 1.40 & \cite{hay96} \\   
HgBa$_2$CuO$_{4+\delta}$              & $P_{pl}$   & 0.235 & 813              & 2.57$\times10^3$ & 3.16 & \cite{yam00} \\
HgBa$_2$CaCu$_2$O$_{6+\delta}$        & $P_{pl}$   & 0.227 & 13.5             & 21.7             & 1.61 & \cite{fuk97} \\
HgBa$_2$Ca$_2$Cu$_3$O$_{8+\delta}$    & $P_{pl}$   & 0.215 & 38.9             & 100              & 2.57 & \cite{fuk97} \\
\hline
\end{tabular}
\end{table*}
 
Now we demonstrate that experimental formula (\ref{form3}) can reproduce the $T$ dependence of $in$-$plane$ resistivity ($\rho_{ab}$) for various HTCSs at different doping levels. In Fig.\ \ref{fig6}, we plot $\rho_{ab}$ for SrD-La214 and Bi$_2$(Sr$_{2-x}$La$_x$)CuO$_{6+\delta}$ (LaD-Bi2201) as a function of $T$ \cite{kom05,and00}. In the inset, we plot resistivity along the $a$-axis ($\rho_a$) and resistivity along the $b$-axis ($\rho_b$) for OD-Y123 \cite{seg03}. In these figures, the solid curves represent the resistivity calculated from formula (\ref{form3}). It can be seen that $\rho_{ab}$ above 110 K for SrD-La214 and LaD-Bi2201, and $\rho_b$ above 120 K for OD-Y123 are well represented by formula (\ref{form3}) with the corresponding value of $\sigma^{opt}(290)$. But, the downward deviation below 140 K in $\rho_a$ of OD-Y123, due to pseudogap effect, cannot be explained by formula (\ref{form3}).     

Here, we will comment on the influence of the pseudogap effect on the $universal$ $intrinsic$ $doping$ $behavior$. As a common feature of the underdoped HTCSs, there is a pseudogap effect at some temperature range above $T_c$. The pseudogap effect in the resistivity measurement is often observed as a deviation from the linear $T$ dependence at high-$T$ or enhancement of conductivity below a characteristic temperature. According to the UEPD, the pseudogap phase appears above 290 K for $p_u < 0.5$ \cite{hon08}. Therefore, the original universal intrinsic doping behavior at 290 K, the formula (\ref{form1}), is not influenced by the pseudogap effect, since it is observed in the $p_u$-range of $0.5 \le p_u < 1$. From the insets of Figs.\ \ref{fig2} and\ \ref{fig4}, the univesal intrinsic doping behavior observed at 290 K in the $p_u$ range of $0.5 \le p_u < 1$ is preserved all the way down to 110 K or $(T_c + 20)$ K within slight scatterings. This shows that even if there is the pseudogap effect in the co-doped HTCS or the Bi-based family, the influence is within the scattering. In fact, the size of the deviation or enhancement of conductivity tends to depend on the material type of HTCSs or sample quality. It is well-known that no clear deviation is observed in SrD-La214 \cite{kom05}. In OD-Y123, the relatively large deviation is observed \cite{seg03,tak94}. The deviation from the linearity in OD-Bi2212 is smaller than that in OD-Y123 \cite{seg03,tak94,wat97}. These suggest that our observed universal intrinsic doping behavior, within some scattering, is preserved in the wide temperature range from $(T_c + 20)$ K to 290 K, where the pseudogap phase appears. Therefore, we conclude that our observed universal intrinsic doping behavior is not significantly influenced by the pseudogap effect. 

\begin{figure}
\includegraphics[scale=0.35]{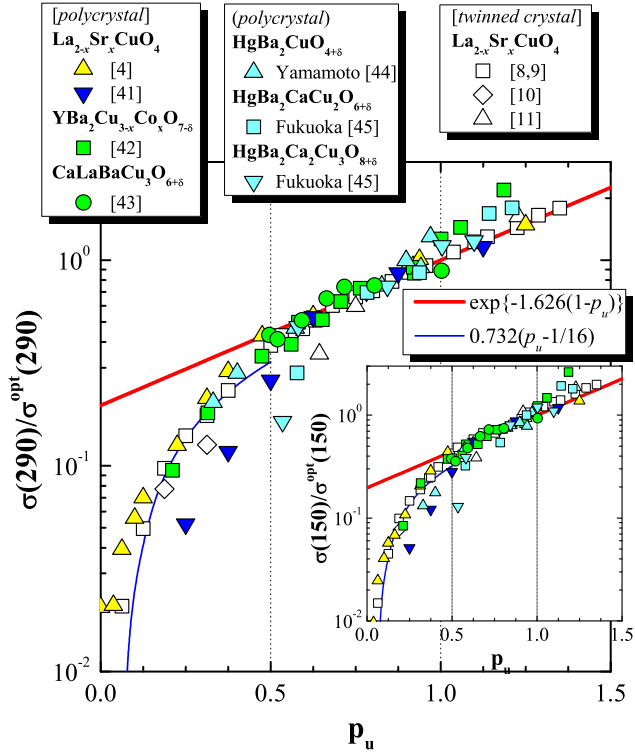}
\caption{\label{fig7}$\sigma (290)/\sigma^{opt}(290)$ versus $p_u$ for the single-crystal and polycrystalline samples. The inset shows the $\sigma(150)/\sigma^{opt}(150)$ versus $p_u$. The sample quality, the method used to determine $P_{pl}$ and the corresponding references of the plotted data are summarized in Table~\ref{tab:table3}.}
\end{figure}

Now let us summarize the intrinsic doping dependence of in-plane conductivity for HTCSs: a linear $p_u$-dependence of dc conductivity appears at a material-dependent critical doped-hole concentration of $p_u \sim$ 1/16, 1/8 or 1/4 and continues up to $p_u \sim 0.5$. Then, it is followed by a universal intrinsic doping behavior, expressed in terms of exponential function, of $in$-$plane$ dc conductivity with increasing $p_u$ for $0.5 \le p_u < 1$. Therefore, the three key findings of our analysis are: (i) a  material-dependent critical doped-hole concentration ($p_u^c$) where metallic-like conductivity appears in the normal state, (ii) a linear $p_u$ dependence of $\sigma$ $\propto$ ($p_u - p_u^c$) for $p_u^c \le p_u < 0.5$, and (iii) an universal intrinsic doping behavior for $0.5 \le p_u < 1$ which can be approximately represented as $\exp\{-1.626(1-p_u)\}$. We should emphasize that the above conclusions are all regulated only by the doped-hole concentration. Only $\sigma^{opt}(290)$, which served as a scaling factor, is dependent on the extended structure and sample quality. While the linear $p_u$ dependence of conductivity in (ii) seems to be quite conventional, (i) and (iii) actually imply a very unusual transport behavior. Actually even (ii) along is unusual because, around room temperature, dc conductivity of SrD-La214 is metallic-like for doping as low as $x$ = 0.01 or $p_u$ = 1/16, which yields a mean free path much smaller than the inter-atomic distance, in clear violation of the Mott limit for metallic conduction \cite{and01}. Accidentally, the existence of $p_u^c \sim$ 1/16, 1/8 and 1/4 in the normal state transport property reminds us of the underling electronic texture due to the mixing of pristine electronic phases at magic doping concentrations $P_{pl} = m/n^2$, where $m$ and $n$ are positive integers with $n$ = 3 or 4 and $m$ $\le$ $n$ \cite{hon07,don08}. For example, in Fig.\ \ref{fig1}(a), the anomaly in the $T_c/T_c^{max}$ versus $p_u$ curve appears at $p_u$ $\sim$ 0.75 = 3/4. The anomalies for OD-Y123 and SrD-La214 are corresponding to the appearance of 90 K plateau and the 1/8 anomaly, respectively. 

Furthermore, we show that even conductivity of polycrystalline samples follow the universal intrinsic doping behavior. Figure\ \ref{fig7} shows $\sigma(290)/\sigma^{opt}(290)$ as a function of $p_u$. We plot $\sigma(290)/\sigma^{opt}(290)$ of single-crystal SrD-La214 \cite{kom05,kom03,xu00,nak93}, and polycrystalline SrD-La214 \cite{tak89,nak94}. We also plot $\sigma(290)/\sigma^{opt}(290)$ of polycrystalline YBa$_2$Cu$_{3-x}$Co$_x$O$_{7-\delta}$ (CoD-Y123), CaLaBaCu$_3$O$_{6+\delta}$ and Hg-based family which also reports the $S^{290}$ value \cite{fis96,yam00,hay96,fuk97}. Since the magnitude of $\sigma(290)$ of polycrystalline samples are consistently lower than that of single crystal, clearly only $\sigma^{opt}(290)$, which is the material-dependent parameter, depends on the preparation conditions. The doping dependence follows formula (\ref{form1}) for $0.5 \le p_u < 1$. Although the doping dependence for $p_u < 0.5$ tends to depend on the samples, it deviates downward from formula (\ref{form1}) below $p_u \sim 0.5$. For $p_u > 1$, the doping dependence in SrD-La214 continues to follow (\ref{form1}), while that in CoD-Y123 deviates upward from it. These features are same as the $in$-$plane$ conductivity for SrD-La214 and Y$_{1-x}$Ca$_x$Ba$_2$Cu$_3$O$_{6+\delta}$ (CaD-Y123). CoD-Y123 also has the disrupted chain like CaD-Y123. Same result is observed also at 150 K as shown in the inset of Fig.\ \ref{fig7}. Although $\sigma^{opt}(150)/\sigma^{opt}(290)$ for HgBa$_2$CuO$_{4+\delta}$ is larger than the other materials, it is noted that $\sigma^{opt}(150)/\sigma^{opt}(290)$ is $\sim$1.8 as shown in Table~\ref{tab:table3}, independent of the materials, and is comparable to that of the single crystal. Therefore we have quantitatively confirmed that the conductivity of the well-characterized polycrystalline samples follows same universal intrinsic doping behavior for $in$-$plane$ dc conductivity, although the magnitude of $\sigma^{opt}$(290) is dependent of the preparation condition for polycrystalline samples.  
\begin{figure}
\includegraphics[scale=0.35]{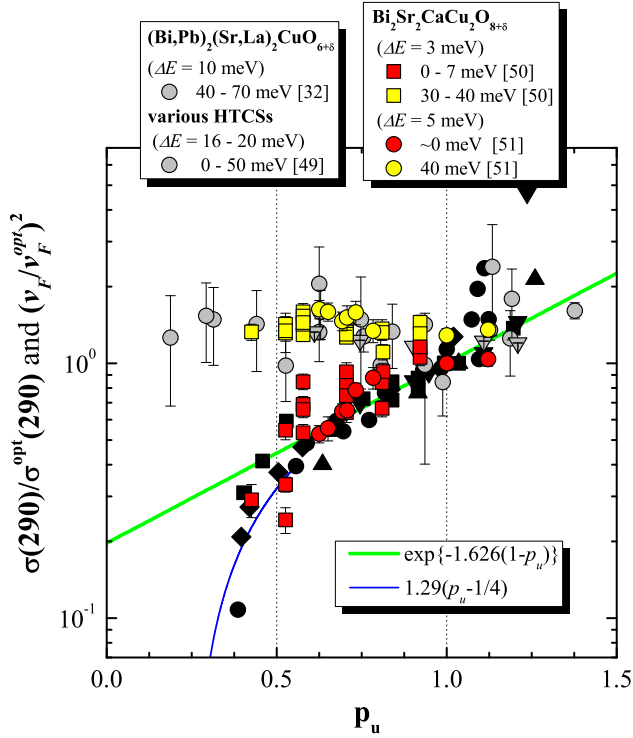}
\caption{\label{fig8}Comparison of ($v_F$/$v_F^{opt}$)$^2$ observed in ARPES with $\sigma$(290)/$\sigma^{opt}$(290) for Bi-based family. The data of normalized conductivity, which are black symbols, are the same as that plotted in Fig.\ \ref{fig4}.}
\end{figure}

The universal intrinsic doping dependence of $in$-$plane$ dc conductivity is anomalous and quite intriguing. We use very generic arguments invoking linear response theory and find that the behavior may come exclusively from the doping dependence of the nodal Fermi velocity ($v_F$). Linear response theory emphasizes that the metallic conductivity is due to the high current carried by the few carriers along the most conducting channel at Fermi level ($E_F$). Quite general, metallic conductivity can be represented as $\sigma \propto j_F^2\tau N(E_F) \propto e^2v_F^2\tau N(E_F)$ \cite{zim72}, where $j_F$ (= $ev_F$), $e$, $\tau$ and $N(E_F)$ are the current density at $E_F$, electron charge, the relaxation time and the density of states at $E_F$, respectively. The $v_F$ value can be extracted from the dispersion of low-energy quasiparticles measured by high-resolution laser-based angle-resolved photoemission spectroscopy (ARPES). In Fig.\ \ref{fig8}, we plot the square nodal $v_F$ normalized to that at the optimal doping level ($v_F^{opt}$) of various HTCSs extrated from Refs.\ \cite{kon06,zho03,vis10,anz10} and $\sigma(290)/\sigma^{opt}(290)$ of Bi-based family plotted in Fig.\ \ref{fig4} as a function of $p_u$ on the semi-logarismic plot. The doping-independent nodal ($v_F$/$v_F^{opt}$)$^2$ extracted from Refs.\ \cite{kon06,zho03} were coming from ARPES measurements at 200 K and 20 K with typical energy resolutions of 10 meV and 16 - 20 meV, respectively. In underdoped OD-Bi2212 at 10 K, a strong doping-dependent nodal $v_F$ is observed in the energy range of 0 - 7 meV through a much improved energy resolution of 3 or 5 meV in laser-based ARPES experiment \cite{vis10,anz10}. The ($v_F$/$v_F^{opt}$)$^2$ values, calculated from the nodal $v_F$ observed in the energy range of 0 - 7 meV, lie on the universal intrinsic doping behavior. Further, the ($v_F$/$v_F^{opt}$)$^2$ values seems to rapidly deviate downward with undoping for $p_u$ $<$ 0.5. The doping dependence of $v_F^2$ is identical to feature (iii) and the deviation is qualitatively similar to feature (ii). Therefore, we are led to the conclusion that the doping dependence of $in$-$plane$ dc conductivity is actually coming from that of the nodal $v_F^2$.
 
Note that the same doping dependence for $v_F^2$ are extracted from two very different experiments for the different temperature regions. One is from dc transport measurements at 290 K or room temperature ($>$ $T_c$) and the other is from high-resolution laser-ARPES at $T$ = 10 K ($<$ $T_c$). It is the first time that one normal state transport property at room temperature is correlated to and exhibits the same doping dependence as the property derived from the superconducting state, namely, transport carriers in normal state are dictated by the similar nodal carriers in the $d$-wave superconductor. This may also imply that the normal state transport is also $d$-wave like, for instance in a $d$-density wave state \cite{cha01}, and the underlying electronic texture, that gave us the $universal$ $intrinsic$ $doping$ $behavior$ of $in$-$plane$ dc conductivity, persists into the superconducting state; both the normal and the superconducting states share a common underlying electronic texture. Furthermore since the universal intrinsic doping behavior is purely electronic in origin which, in turn, suggests a universal $p_u$ dependence of nodal $v_F$ for HTCSs. 

Finally, we want to comment, independently of present study, on our study of the doping dependence of $out$-$of$-$plane$ ($c$-axis) conductivity for the hole-doped underdoped HTCSs \cite{hon10}. Although the analyzing method based on our hole-scale is same, the doping dependence of $out$-$of$-$plane$ conductivity is characteristically different from that of $in$-$plane$ conductivity. Specifically, $c$-axis conductivity has $c$-direction contributions of a doping-dependent-activated gap and a quantum tunneling between adjacent CuO$_2$ layers sandwiched by the insulator-like block layers \cite{hon10}. The differences in conductivities along $out$-$of$-$plane$ and $in$-$plane$ directions is one of the most puzzling issues in transport properties of HTCSs which is closely related to the detail mechanism of conduction and the doping evolution of the electronic structure along each directions. We hope our work will stimulate further theoretical efforts to address the microscopic understanding of the universal doping dependence along $ab$-palne and $c$-axis in the underdoped cuprate superconductors.

\section{Conclusions}

Using our $P_{pl}$-scale or $p_u$-scale and the single crystal data accumulated over the past twenty years, we have revealed the hidden universal intrinsic $in$-$plane$ dc conductivity behavior for hole-doped HTCSs. $In$-$plane$ dc conductivity exhibits a linear $p_u$ dependence of $\sigma \propto (p_u - p_u^c)$ for $p_u^c \le p_u < 0.5$ ($p_u^c$ $\sim$ 1/16, 1/8 or 1/4) and an nominal exponential $p_u$ dependence of $\sigma \propto \exp\{-1.626(1-p_u)\}$ for $0.5 \le p_u < 1$. This $universal$ $intrinsic$ $doping$ $behavior$ is consistent with the doping dependence of the nodal $v_F$ derived in the superconducting state. Our findings suggest a commonality of the low-energy quasiparticles both in the normal and superconducting electronic states of normal carriers that place a true universal and stringent constraint on the mechanism of high-$T_c$ superconductivity of HTCSs. 

\section*{Acknowledgements}

PHH acknowledges the support of the State of Texas through the Texas Center for Superconductivity at the University of Houston.


\begin{thebibliography}{00}

\bibitem{hon04} T. Honma, P. H. Hor, H. H. Hsieh and M. Tanimoto, Phys. Rev. B 70 (2004) 214517. 
\bibitem{hon08} T. Honma and P. H. Hor, Phys. Rev. B 77 (2008) 184520.
\bibitem{pre91} M. R. Presland et. al., Physica C 176 (1991) 95.
\bibitem{tak89} H. Takagi et. al., Phys. Rev. B 40 (1989) 2254.
\bibitem{rad94} For example, P. G. Radaelli et. al., Phys. Rev. B 49 (1994) 4163.
\bibitem{hon07} T. Honma and P. H. Hor, Phys. Rev. B 75 (2007) 012508.
\bibitem{hon06} T. Honma and P. H. Hor, Supercond. Sci. Tech. 19 (2006) 907. 
\bibitem{kom05} S. Komiya, H. -D. Chen, S. -C. Zhang and Y. Ando, Phys. Rev. Lett. 94 (2005) 207004. 
\bibitem{kom03} S. Komiya, X. F. Sun, A. N. Lavrov and Y. Ando, Physica C 392-396 (2003) 135.
\bibitem{xu00}  Z. A. Xu et. al., Physica C 341-348 (2000) 1711.
\bibitem{nak93} Y. Nakamura and S. Uchida, Phys. Rev. B 47 (1993) 8369.
\bibitem{abe99} Y. Abe et. al., Phys. Rev. B 59 (1999) 14753.
\bibitem{hay91} S. M. Hayden et al., Phys. Rev. Lett. 66 (1991) 821.
\bibitem{nak92} Y. Nakamura and S. Uchida, Phys. Rev. B 46(1992) 5841.
\bibitem{wak04} K. Waku et. al., Phys. Rev. B 70 (2004) 134501.
\bibitem{koh03} Y. Kohsaka et. al., J. Phys. Soc. Jpn. 72 (2003) 1018. 
\bibitem{dai01} A. Daignere, A. Wahl, V. Hardy and A. Maignan, Physica C 349 (2001) 189. 
\bibitem{nag08} K. Nagasao, T. Matsui and S. Tajima, Physica C 468 (2008) 1188. 
\bibitem{yak95} H. Yakabe et. al., Jpn. J. Appl. Phys. 34 (1995) 4754. 
\bibitem{noj03} T. Noji, H. Akagawa, Y. Ono and Y. Koike, J. Low Temp. Phys. 131 (2003) 699.
\bibitem{seg03} K. Segawa and Y. Ando, J. Low Temp. Phys. 131 (2003) 821.
\bibitem{tak94} K. Takenaka, K. Mizuhashi, H. Takagi and S. Uchida, Phys. Rev. B 50 (1994) 6534. 
\bibitem{lee05} Y. S. Lee et. al., Phys. Rev. B 72 (2005) 054529. 
\bibitem{coo00} J. R. Cooper et. al., Physica C 341-348 (2000) 855. 
\bibitem{bab99} D. Babi\'{c}, J. R. Cooper, J. W. Hodby and C. Changkang, Phys. Rev. B 60 (1999) 698. 
\bibitem{sem01} K. Semba and A. Matsuda, Phys. Rev. Lett. 86 (2001) 496.
\bibitem{zve03} V. N. Zverev and D. V. Shovkun, Physica C 391 (2003) 315. 
\bibitem{wan01} Y. Wang and N. P. Ong, Proc. Nat. Acad. Sci. 98 (2001) 11091.
\bibitem{and00} Y. Ando et. al., Phys. Rev. B 61 (2000) R14956.
\bibitem{men09} J. Meng et. al., Supercond. Sci. Tech. 22 (2009) 045010. 
\bibitem{ama04} T. Amano et. al., Physica C 412-414 (2004) 230. 
\bibitem{kon06} T. Kondo, T. Takeuchi, S. Tsuda and S. Shin, Phys. Rev. B 74 (2006) 224511. 
\bibitem{fuj02} T. Fujii, I. Terasaki, T. Watanabe and A. Matsuda, Physica C 378-381 (2002) 182. 
\bibitem{wat97} T. Watanabe, T. Fujii and A. Matsuda, Phys. Rev. Lett. 79 (1997) 2113. 
\bibitem{fuj95} A. Fujiwara et. al., Phys. Rev. B 52 (1995) 15598. 
\bibitem{yam07} Y. Yamada, T. Watanabe and M. Suzuki, Physica C 460-462 (2007) 815.
\bibitem{fuj02b} T. Fujii, I. Terasaki, T. Watanabe and A. Matsuda, Phys. Rev. B 66 (2002) 024507. 
\bibitem{iye92} Y. Iye , J. Phys. Chem. Solids 53 (1992) 1561.
\bibitem{yam90} A. Yamamoto et. al., Phys. Rev. B 42 (1990) 4228.
\bibitem{hon91} T. Honma, K. Yamaya, F. Minami and S. Takekawa, Physica C 176 (1991) 209.
\bibitem{nak94} T. Nakano et. al., Phys. Rev. B 49 (1994) 16000.
\bibitem{fis96} B. Fisher et. al., J. Appl. Phys. 80 (1996) 898.
\bibitem{hay96} K. Hayashi et. al., Czec. J. Phys. 46(S2) (1996) 1181.
\bibitem{yam00} A. Yamamoto, W. -Z. Hu and S. Tajima, Phys. Rev. B 63 (2000) 024504.
\bibitem{fuk97} A. Fukuoka et. al., Phys. Rev. B 55 (1997) 6612. 
\bibitem{and01} Y. Ando et. al., Phys. Rev. Lett. 87 (2001) 017001.
\bibitem{don08} X. L. Dong, P. H. Hor, F. Zhou and Z. -X. Zhao, Solid State Commun. 145 (2008) 173. 
\bibitem{zim72} J. M. Zimann, Principles of the theory of solids, 2nd ed., Cambridge Univ. Press., Cambridge, 1972. 
\bibitem{zho03} X. J. Zhou et. al.,Nature (London) 423 (2003) 398.
\bibitem{vis10} I. M. Vishik et. al., Phys. Rev. Lett. 104 (2010) 207002.
\bibitem{anz10} H. Anzai et. al., arXiv:1004.3961v1. 
\bibitem{cha01} S. Chakravarty, R. B. Laughlin, D. K. Morr and C. Nayak, Phys. Rev. B 63 (2001) 094503. 
\bibitem{hon10} T. Honma and P. H. Hor, Solid State Commun. 150 (2010) 10857.

\end{thebibliography}
\end{document}